\documentclass[aps,prb,twocolumn,showpacs,floatfix]{revtex4}

\usepackage{graphicx}

\begin{document}

\title{One-loop approximation for the Hubbard model}

\author{A.~Sherman}

\affiliation{Institute of Physics, University of Tartu, Riia 142, 51014
Tartu, Estonia}

\date{\today}

\begin{abstract}
The diagram technique for the one-band Hubbard model is formulated for
the case of moderate to strong Hubbard repulsion. The expansion in
powers of the hopping constant is expressed in terms of site cumulants
of electron creation and annihilation operators. For Green's function
an equation of the Larkin type is derived and solved in a one-loop
approximation for the case of two dimensions, nearest-neighbor hopping
and half-filling. The obtained four-band structure of the spectrum and
the shapes of the spectral function are close to those observed in
Monte Carlo calculations. It is shown that the maxima forming the bands
are of a dissimilar origin in different regions of the Brillouin zone.
\end{abstract}


\maketitle

\section{Introduction}
Systems with strong electron correlations which, in particular, cuprate
perovskites belong to are characterized by a Coulomb repulsion that is
comparable to or larger than hopping constants. One of the simplest
models for the description of such systems is the Hubbard model
\cite{Gutzwiller,Hubbard63,Kanamori} with the Hamiltonian
\begin{equation}\label{Hamiltonian}
H=\sum_{\bf nm\sigma}t_{\bf nm}a^\dagger_{\bf n\sigma}a_{\bf m\sigma} +
\frac{U}{2}\sum_{\bf n\sigma}n_{\bf n\sigma}n_{\bf n,-\sigma},
\end{equation}
where $t_{\bf nm}$ is the hopping constants, the operator
$a^\dagger_{\bf n\sigma}$ creates an electron on the site {\bf n} with
the spin projection $\sigma=\pm 1$, $U$ is the on-site Coulomb
repulsion, the electron number operator $n_{\bf n\sigma}=a^\dagger_{\bf
n\sigma}a_{\bf n\sigma}$. In the case of strong electron correlations,
$U\agt|t_{\bf nm}|$, it is reasonable to use a perturbation expansion
in powers of the hopping constants for investigating Hamiltonian
(\ref{Hamiltonian}). Apparently the first such expansion was considered
in Ref.~\onlinecite{Hubbard66}. The further development of this
approach was given in
Refs.~\onlinecite{Westwanski,Stasyuk,Zaitsev,Izyumov,Ovchinnikov} where
the diagram technique for Hubbard operators was developed and used for
investigating the Mott transition, the magnetic phase diagram and the
superconducting transition in the Hubbard model.

The rules of the diagram technique for Hubbard operators are rather
intricate. Besides, these rules and the graphical representation of the
expansion vary depending on the choice of the operator precedence. The
diagram technique suggested in
Refs.~\onlinecite{Vladimir,Metzner,Moskalenko,Pairault} is free from
these defects. In this approach the power expansion for Green's
function of electron operators $a_{\bf n\sigma}$ and $a^\dagger_{\bf
n\sigma}$ is considered and the terms of the expansion are expressed as
site cumulants of these operators. Based on this diagram technique the
equations of the Larkin type \cite{Larkin} for Green's function were
derived. \cite{Vladimir,Moskalenko,Pairault}

However, the application of this approach runs into problems. In
particular, at half-filling the spectral weight appears to be negative
near frequencies $\omega_d=\pm\frac{U}{2}$. \cite{Pairault} This
drawback is connected with divergencies in cumulants at these
frequencies. As can be seen from formulas given below, all higher-order
cumulants have such divergencies at $\omega_d$ with sign-changing
residues, which are expected to compensate each other in the entire
series. If, as in Ref.~\onlinecite{Pairault}, only some subset of
diagrams is taken into account the divergencies of different orders can
be compensated only accidentally. On the other hand, at frequencies
neighboring to $\omega_d$ cumulants are regular. If a selected subset
of diagrams is expected to give a correct estimate of the entire series
for these frequencies the values at $\omega_d$ can be corrected by an
interpolation using results for the regular regions. This procedure is
applied in the present work where the one-loop approximation is used --
as irreducible diagrams in Larkin's equation those appearing in the
first two orders of the perturbation expansion are adopted. One of
these diagrams contains a loop formed by a hopping line. It is shown
that for the case of a square lattice, nearest-neighbor hopping and
half-filling the electron spectrum consists of four bands. These band
structure and the calculated shapes of the electron spectral function
are close to those obtained in Monte-Carlo calculations
\cite{Moreo,Preuss,Grober} and recently in the cluster perturbation
\cite{Dahnken} and the two-particle self-consistent \cite{Tremblay}
calculations. It is shown also that the maxima forming the bands are of
a dissimilar origin in different regions of the Brillouin zone.

In the following section the perturbation expansion for the electron
Green's function is formulated in the form convenient for calculations
and Larkin's equation is derived. In Sec.~III the equations of the
previous section are used for calculating the spectral function and the
obtained results are discussed. Concluding remarks are presented in
Sec.~IV.

\section{Diagram technique}
We consider Green's function
\begin{equation}\label{GF}
G({\bf n\tau,n'\tau'})=\langle{\cal T}\bar{a}_{\bf n'\sigma}(\tau')
a_{\bf n\sigma}(\tau)\rangle,
\end{equation}
where the angular brackets denote the statistical averaging with the
Hamiltonian ${\cal H}=H-\mu\sum_{\bf n\sigma}n_{\bf n\sigma}$, $\mu$ is
the chemical potential, ${\cal T}$ is the time-ordering operator which
arranges other operators from right to left in ascending order of times
$\tau$, $a_{\bf n\sigma}(\tau)=\exp({\cal H}\tau)a_{\bf
n\sigma}\exp(-{\cal H}\tau)$ and $\bar{a}_{\bf
n\sigma}(\tau)=\exp({\cal H}\tau)a^\dagger_{\bf n\sigma}\exp(-{\cal
H}\tau)$. Choosing
\begin{eqnarray}
H_0&=&\frac{U}{2}\sum_{\bf n\sigma}n_{\bf n\sigma}n_{\bf
n,-\sigma}-\mu\sum_{\bf
 n\sigma}n_{\bf n\sigma},\nonumber\\[-1ex]
&&\label{division}\\[-1ex]
H_1&=&\sum_{\bf nm\sigma}t_{\bf nm}a^\dagger_{\bf n\sigma}a_{\bf
 m\sigma}\nonumber
\end{eqnarray}
as the unperturbed Hamiltonian and the perturbation, respectively, and
using the known \cite{Abrikosov} expansion for the evolution operator
we get
\begin{eqnarray}
G({\bf n'\tau',n\tau})&=&\sum_{k=0}^\infty \frac{(-1)^k}{k!}
 \int\!\ldots\!\int_0^\beta d\tau_1\ldots d\tau_k\nonumber\\
&\times&\sum_{{\bf n}_1{\bf n}^\prime_1\sigma_1}\!\!\ldots\!\!
 \sum_{{\bf n}_k{\bf n}^\prime_k\sigma_k} t_{{\bf n}_1{\bf n}^\prime_1}
 \ldots t_{{\bf n}_k{\bf n}^\prime_k}\nonumber\\
&\times&\langle{\cal T}\bar{a}_{\bf n'\sigma}(\tau') a_{\bf
 n\sigma}(\tau) \bar{a}_{{\bf n}^\prime_1\sigma_1}(\tau_1) a_{{\bf
 n}_1\sigma_1}(\tau_1)\ldots\nonumber\\[1ex]
&\times&\bar{a}_{{\bf n}^\prime_k\sigma_k}(\tau_k)
 a_{{\bf n}_k\sigma_k}(\tau_k)\rangle_{0c}, \label{series}
\end{eqnarray}
where $\beta=1/T$ is the inverse temperature, the subscript ``0'' near
the angular bracket indicates that the averaging and time dependencies
of operators are determined with the Hamiltonian $H_0$. The subscript
``c'' indicates that terms which split into two or more disconnected
averages have to be dropped out.

The Hamiltonian $H_0$ is diagonal in the site representation. Therefore
the average in the right-hand side of Eq.~(\ref{series}) splits into
averages belonging to separate lattice sites. To be nonzero these
latter averages have to contain equal number of creation and
annihilation operators. Let us consider the average which appears in
the first order: $\langle{\cal T}\bar{a}_{\bf n'\sigma}(\tau') a_{\bf
 n\sigma}(\tau) \bar{a}_{{\bf n}^\prime_1\sigma_1}(\tau_1) a_{{\bf
 n}_1\sigma_1}(\tau_1)\rangle$ (for short here and below subscripts
``0'' and ``c'' are omitted). For this average to be nonvanishing
operators have to belong either to the same site or to two different
sites,
\begin{eqnarray*}
&&\langle{\cal T}\bar{a}_{\bf n'\sigma}(\tau') a_{\bf
 n\sigma}(\tau) \bar{a}_{{\bf n}^\prime_1\sigma_1}(\tau_1^\prime)
 a_{{\bf n}_1\sigma_1}(\tau_1)\rangle=\\
&&\quad\langle{\cal T}\bar{a}_{\bf n\sigma}(\tau')
 a_{\bf n\sigma}(\tau) \bar{a}_{{\bf n}\sigma_1}(\tau_1^\prime)
 a_{{\bf  n}\sigma_1}(\tau_1)\rangle \delta_{\bf nn'}
 \delta_{{\bf nn}_1^\prime} \delta_{{\bf nn}_1}\\
&&\quad+\langle{\cal T}\bar{a}_{\bf n\sigma}(\tau') a_{\bf
 n\sigma}(\tau)\rangle \langle{\cal T}\bar{a}_{{\bf n}_1\sigma_1}
 (\tau_1^\prime) a_{{\bf n}_1\sigma_1}(\tau_1)\rangle\\
&&\quad\quad\times\delta_{\bf nn'}
 \delta_{{\bf n}_1{\bf n}_1^\prime} (1-\delta_{{\bf nn}_1})\\
&&\quad-\langle{\cal T}\bar{a}_{{\bf n}_1\sigma}(\tau') a_{{\bf
 n}_1\sigma_1}(\tau_1)\rangle \langle{\cal T}\bar{a}_{{\bf n}\sigma_1}
 (\tau_1^\prime) a_{\bf n\sigma}(\tau)\rangle\\
&&\quad\quad\times\delta_{{\bf n'n}_1}
 \delta_{{\bf nn}_1^\prime} (1-\delta_{{\bf nn}_1}).
\end{eqnarray*}
The multiplier $1-\delta_{{\bf nn}_1}$ ensures that ${\bf n}\neq{\bf
n}_1$ in the two last terms (the case ${\bf n}={\bf n}_1$ is taken into
account by the first term in the right-hand side). After the
rearrangement of the terms we find
\begin{eqnarray*}
&&\langle{\cal T}\bar{a}_{\bf n'\sigma}(\tau') a_{\bf
 n\sigma}(\tau) \bar{a}_{{\bf n}^\prime_1\sigma_1}(\tau_1^\prime)
 a_{{\bf n}_1\sigma_1}(\tau_1)\rangle=\\
&&\quad K_2(\tau'\sigma,\tau\sigma,\tau'_1\sigma_1,\tau_1\sigma_1)
 \delta_{\bf nn'} \delta_{{\bf nn}_1^\prime} \delta_{{\bf nn}_1}\\
&&\quad+K_1(\tau'\sigma,\tau\sigma)K_1(\tau'_1\sigma_1,\tau_1\sigma_1)
 \delta_{\bf nn'} \delta_{{\bf n}_1{\bf n}'_1}\\
&&\quad-K_1(\tau'\sigma,\tau_1\sigma_1)K_1(\tau'_1\sigma_1,\tau\sigma)
 \delta_{{\bf n'n}_1} \delta_{{\bf nn}'_1},
\end{eqnarray*}
where
\begin{eqnarray}
&&K_1(\tau'\sigma',\tau\sigma)=\langle{\cal T}\bar{a}_\sigma(\tau')
 a_\sigma(\tau)\rangle \delta_{\sigma\sigma'},\nonumber\\
&&K_2(\tau'\sigma,\tau\sigma,\tau'_1\sigma_1,\tau_1\sigma_1)=\nonumber\\
&&\quad\langle{\cal T}\bar{a}_\sigma(\tau') a_\sigma(\tau)
 \bar{a}_{\sigma_1}(\tau'_1) a_{\sigma_1}(\tau_1)\rangle
 \label{cumulants}\\
&&\quad -K_1(\tau'\sigma,\tau\sigma)K_1(\tau'_1\sigma_1,\tau_1\sigma_1)
 \nonumber\\
&&\quad +K_1(\tau'\sigma,\tau_1\sigma_1)K_1(\tau'_1\sigma_1,\tau\sigma)
 \nonumber
\end{eqnarray}
are cumulants\cite{Kubo} of the first and second order, respectively.
All operators in cumulants~(\ref{cumulants}) belong to the same lattice
site. Due to the translational symmetry of $H_0$ in
Eq.~(\ref{division}) the cumulants do not depend on the site index
which is therefore omitted in Eq.~(\ref{cumulants}).

\begin{figure}[t]
\includegraphics[width=8.5cm]{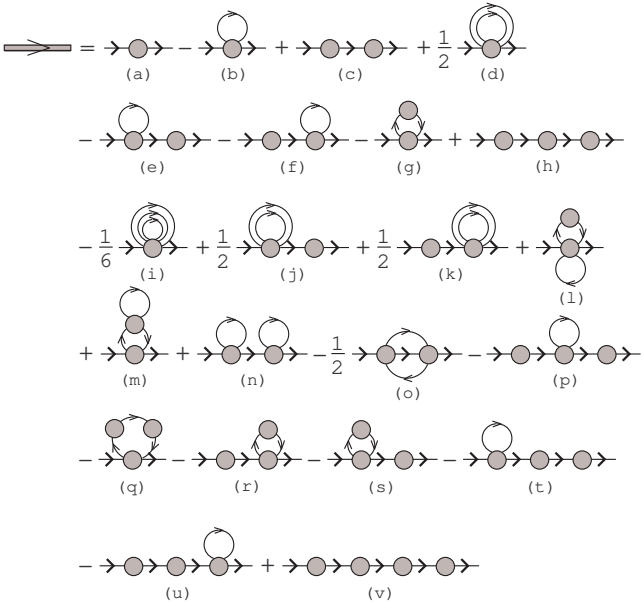}
\caption{Diagrams of the first four orders of expansion
(\protect\ref{series}).}\label{Fig_i}
\end{figure}

Averages which appear in higher-order terms of Eq.~(\ref{series}) can
be transformed in the same manner. The average in the $k$-th order term
which contains $k+1$ creation and annihilation operators is represented
by the sum of all possible products of cumulants with the sum of orders
equal to $k+1$. All possible distributions of operators between the
cumulants in these products have to be taken into account. The sign of
a term in this sum is determined by the number of permutations of
fermion operators, which bring the sequence of operators in the initial
average to that in the term. Actually the above statements determine
rules of the diagram technique. Additionally we have to take into
account the presence of topologically equivalent terms -- terms which
differ only by permutation of operators $H_1(\tau_i)$ in
Eq.~(\ref{series}). Since these terms are equal, in the expansion only
one of them can be taken into account with the prefactor $\nu=j/k!$
where $j$ is the number of topologically equivalent terms. Following
Ref.~\onlinecite{Metzner} in diagrams we denote a cumulant by a circle
and the hopping constant $t_{\bf nn'}$ by a line directed from ${\bf
n'}$ to ${\bf n}$. The external operators $\bar{a}_{\bf
n'\sigma}(\tau')$ and $a_{\bf n\sigma}(\tau)$ are denoted by directed
lines leaving from and entering into a cumulant. The order of a
cumulant is equal to a number of incoming or outgoing lines. Summations
and integrations over the internal indices ${\bf n}_i$, ${\bf n}'_i$,
$\sigma_i$ and $\tau_i$ are carried out. Since site indices of
operators included in a cumulant coincide, some site summations
disappear. Also some summations over $\sigma_i$ get lost, because in
any cumulant spin indices of creation and annihilation operators have
to match. Taking into account the multiplier $(-1)^k$ in
Eq.~(\ref{series}), the sign of the diagram is equal to $(-1)^l$ where
$l$ is the number of loops formed by hopping lines. Figure~\ref{Fig_i}
demonstrates connected diagrams of the first four orders of the power
expansion~(\ref{series}) with their signs and prefactors. Here the
thick line with arrow in the left-hand side of the equation is the
total Green's function. Notice that if we set $t_{\bf nn}=0$,
contributions of the diagrams (b), (d)--(f), (i)--(n), (p), (t), and
(u) vanish. However, below a renormalized hopping parameter will be
introduced which is nonzero for coinciding site indices and therefore
the mentioned diagrams are retained in Fig.~\ref{Fig_i}.

All diagrams can be separated into two categories -- those which can be
divided into two parts by cutting some hopping line and those which
cannot be divided in this way.\cite{Larkin,Izyumov} These latter
diagrams are referred to as irreducible diagrams. In Fig.~\ref{Fig_i}
the diagrams (c), (e), (f), (h), (j), (k), (n), (p), and (r)--(v)
belong to the first category, while the others are from the second
category. The former diagrams can be constructed from the irreducible
diagrams by connecting them with hopping lines. Notice that a prefactor
$\nu$ of some composite diagram is the product of prefactors of its
irreducible parts [cf., e.g., the irreducible diagrams (a), (d) and the
composed diagrams (j), (k) in Fig.~\ref{Fig_i}]. If we denote the sum
of all irreducible diagrams as $K({\bf n'}\tau',{\bf n}\tau)$, the
equation for Green's function can be written as
\begin{eqnarray}
G({\bf n'}\tau',{\bf n}\tau)&=&K({\bf n'}\tau',{\bf n}\tau)+
 \sum_{{\bf n}_1{\bf n}'_1} \int_0^\beta d\tau_1 K({\bf
 n'}\tau',{\bf n}_1\tau_1)\nonumber\\
&\times&t_{{\bf n}_1{\bf n}'_1} G({\bf n}'_1\tau_1,{\bf
n}\tau).\label{Larkin}
\end{eqnarray}
This equation has the form of the Larkin equation. \cite{Larkin}

The partial summation can be carried out in the hopping lines of
cumulants by inserting the irreducible diagrams into these lines. In
doing so the hopping constant $t_{\bf nn'}$ in the respective formulas
is substituted by
\begin{eqnarray}
\Theta({\bf n\tau,n'\tau'})&=&t_{\bf nn'}\delta(\tau-\tau')
 \nonumber\\
&+&\sum_{{\bf n}_1{\bf n}'_1}t_{{\bf nn}'_1}\int_0^\beta d\tau_1
 K({\bf n}'_1\tau,{\bf n}_1\tau_1)\nonumber\\
&\times&\Theta({\bf n}_1\tau_1,{\bf n}'\tau'). \label{hopping}
\end{eqnarray}
For the diagram (b) in Fig.~\ref{Fig_i} inserting irreducible diagrams
into the hopping line leads to the diagrams (g), (m), and (q).

Due to the translation and time invariance of the problem the
quantities in Eqs.~(\ref{Larkin}) and~(\ref{hopping}) depend only on
differences of site indices and times. The use of the Fourier
transformation
$$G({\bf k},i\omega_l)=\sum_{\bf n'}e^{i{\bf k(n'-n)}}\int_0^\beta
e^{i\omega_l(\tau'-\tau)}G({\bf n}'\tau',{\bf n}\tau),$$ where
$\omega_l=(2l+1)\pi T$ is the Matsubara frequency, simplifies
significantly these equations:
\begin{eqnarray}
G({\bf k},i\omega_l)&=&\frac{K({\bf k},i\omega_l)}{1-t_{\bf k}K({\bf
 k},i\omega_l)},\label{tLarkin}\\
\Theta({\bf k},i\omega_l)&=&\frac{t_{\bf k}}{1-t_{\bf k}K({\bf
 k},i\omega_l)}\nonumber\\[1ex]
&=&t_{\bf k}+t_{\bf k}^2G({\bf k},i\omega_l).\label{thopping}
\end{eqnarray}

In the approximation used below the total collection of irreducible
diagrams $K({\bf k},i\omega_l)$ is substituted by the sum of the two
simplest diagrams (a) and (b) appearing in the first two orders of the
perturbation theory. Due to the form of the latter diagram this
approximation is referred to as the one-loop approximation. In the
diagram (b) the hopping line is renormalized in accordance with
Eq.~(\ref{thopping}). Thus,
\begin{eqnarray}
K(i\omega_l)&=&K_1(i\omega_l)
  -T\sum_{l_1\sigma_1}K_2(i\omega_l\sigma,i\omega_{l_1}\sigma_1,
  i\omega_{l_1}\sigma_1)\nonumber\\
&\times&\frac{1}{N}\sum_{\bf k}t^2_{\bf k}G({\bf
  k},i\omega_{l_1}),\label{approx}
\end{eqnarray}
where $K_1(i\omega_l)$ and
\begin{eqnarray*}
&&K_2(i\omega_{l'}\sigma,i\omega_l\sigma,i\omega_{l'_1}\sigma_1,
i\omega_{l_1}\sigma_1)\\
&&\quad=\int\!\!\!\!\int\!\!\!\!\int\!\!\!\!\int_0^\beta d\tau' d\tau
 d\tau'_1 d\tau_1 e^{-i\omega_{l'}\tau'+i\omega_l\tau-i\omega_{l'_1}
 \tau'_1+i\omega_{l_1}\tau_1}\\[0.5ex]
&&\quad\quad\times K_2(\tau'\sigma,\tau\sigma,\tau'_1\sigma_1,
 \tau_1\sigma_1)\\[1ex]
&&\quad=\beta\delta_{l+l_1,l'+l'_1}K_2(i\omega_l\sigma,i\omega_{l'_1}
 \sigma_1, i\omega_{l_1}\sigma_1)
\end{eqnarray*}
are the Fourier transforms of cumulants~(\ref{cumulants}), $N$ is the
number of sites and we set $\sum_{\bf k}t_{\bf k}=0$. Notice that in
this approximation $K$ does not depend on momentum.

Now we need to calculate the cumulants in Eq.~(\ref{approx}). To do
this it is convenient to introduce the Hubbard operators $X^{ij}_{\bf
n}=|i{\bf n}\rangle\langle j{\bf n}|$ where $|i{\bf n}\rangle$ are
eigenvectors of site Hamiltonians forming $H_0$, Eq.~(\ref{division}).
For each site there are four such states: the empty state $|0{\bf
n}\rangle$ with the energy $E_0=0$, the two degenerate singly occupied
states $|\sigma{\bf n}\rangle$ with the energy $E_1=-\mu$ and the
doubly occupied state $|2{\bf n}\rangle$ with the energy $E_2=U-2\mu$.
The Hubbard operators are connected by the relations
\begin{equation}\label{Hubbard}
a_{\bf n\sigma}=X^{0\sigma}_{\bf n}+\sigma X^{-\sigma,2}_{\bf n},\quad
a^\dagger_{\bf n\sigma}=X^{\sigma 0}_{\bf n}+\sigma X^{2,-\sigma}_{\bf
n}
\end{equation}
with the creation and annihilation operators. The commutation relations
for the Hubbard operators are easily derived from their definition.
Using Eq.~(\ref{Hubbard}) the first cumulant in Eq.~(\ref{cumulants})
can be computed straightforwardly:
\begin{equation}\label{firstc}
K_1(i\omega_l)=\frac{1}{Z_0}\left(\frac{e^{-\beta E_\sigma}+e^{-\beta
E_0}}{i\omega_l-E_{\sigma0}}+\frac{e^{-\beta E_2}+e^{-\beta
E_\sigma}}{i\omega_l-E_{2\sigma}}\right),
\end{equation}
where $Z_0=e^{-\beta E_0}+2e^{-\beta E_\sigma}+e^{-\beta E_2}$ is the
partition function, $E_{ij}=E_i-E_j$. As indicated in
Refs.~\onlinecite{Izyumov,Vladimir,Pairault}, if $K({\bf k},i\omega_l)$
is approximated by this cumulant the result corresponds to the
Hubbard-I approximation. \cite{Hubbard63}

To calculate $K_2$ it is convenient to use Wick's theorem for Hubbard
operators: \cite{Westwanski,Stasyuk,Zaitsev,Izyumov}
\begin{eqnarray}
&&\langle{\cal T}X_{\alpha_1}(\tau_1)\ldots X_{\alpha_i}(\tau_i)
 X_\alpha(\tau)X_{\alpha_{i+1}}(\tau_{i+1})\ldots
 X_{\alpha_n}(\tau_n)\rangle\nonumber\\
&&\quad=\sum_{k=1}^n (-1)^{P_k} g_\alpha(\tau-\tau_k)
 \langle{\cal T}X_{\alpha_1}(\tau_1)\ldots\nonumber\\
&&\quad\quad\times[X_{\alpha_k},X_\alpha]_\pm(\tau_k)\ldots
 X_{\alpha_n}(\tau_n)\rangle, \label{Wick}
\end{eqnarray}
where the averaging and time dependencies of operators are determined
by the Hamiltonian $H_0$, $\alpha$ is the index combining the state and
site indices of the Hubbard operator. If $X_\alpha$ is a fermion
operator ($X^{0\sigma}$, $X^{\sigma2}$ and their conjugates), $P_k$ is
the number of permutation with other fermion operators which is
necessary to transfer the operator $X_\alpha$ from its position in the
left-hand side of Eq.~(\ref{Wick}) to the position in the right-hand
side. In this case
\begin{equation}\label{fermion}
g_\alpha(\tau)=\frac{e^{E_{ij}\tau}}{e^{\beta E_{ij}}+1}\left\{
\begin{array}{rl}-1, & \tau>0, \\ e^{\beta E_{ij}}, & \tau<0,
\end{array} \right.
\end{equation}
where $i$ and $j$ are the state indices of $X_\alpha$. If $X_\alpha$ is
a boson operator ($X^{00}$, $X^{22}$, $X^{\sigma\sigma'}$, $X^{02}$,
and $X^{20}$), $P_k=0$ and
\begin{equation}\label{boson}
g_\alpha(\tau)=\frac{e^{E_{ij}\tau}}{e^{\beta E_{ij}}-1}\left\{
\begin{array}{rl}1, & \tau>0, \\ e^{\beta E_{ij}}, & \tau<0.
\end{array} \right.
\end{equation}
In Eq.~(\ref{Wick}), $[X_{\alpha_k},X_\alpha]_\pm$ denotes an
anticommutator when both operators are of fermion type and a commutator
in other cases.

The substitution of Eq.~(\ref{Hubbard}) in $\langle{\cal T}
\bar{a}_\sigma (\tau')a_\sigma(\tau)\bar{a}_{\sigma_1}(\tau'_1)$
$\times a_{\sigma_1}(\tau_1)\rangle$ in $K_2$, Eq.~(\ref{cumulants}),
leads to six nonvanishing averages of Hubbard operators (such averages
are nonzero if the numbers of the operators $X^{0\sigma}$ and
$X^{\sigma 0}$ coincide in them, and the same is true for the pair
$X^{\sigma 2}$ and $X^{2\sigma}$). Applying Wick's theorem (\ref{Wick})
to these averages the number of operators in them is sequentially
decreased until only time-independent operators are retained. For $H_0$
in Eq.~(\ref{division}) these are $X^{00}$, $X^{\sigma\sigma'}$, and
$X^{22}$. Their averages are easily calculated. As a result, after some
algebra we find
\begin{eqnarray}
&&\sum_{\sigma_1}K_2(i\omega_l\sigma,i\omega_{l_1}\sigma_1,
 \omega_{l_1}\sigma_1)=-Z_0^{-1}U \bigl\{e^{-\beta E_0}g_{0\sigma}
 (i\omega_l)\nonumber \\
&&\quad\times g_{0\sigma}(i\omega_{l_1})g_{02}(i\omega_l+
 i\omega_{l_1})\bigl[g_{0\sigma}(i\omega_l)+g_{0\sigma}
 (i\omega_{l_1})\bigr] \nonumber\\
&&\quad+e^{-\beta E_2}g_{\sigma2}(i\omega_l)g_{\sigma2}
 (i\omega_{l_1})g_{02}(i\omega_l+i\omega_{l_1})\nonumber\\
&&\quad\times\bigl[g_{\sigma2}(i\omega_l)+g_{\sigma2}(i\omega_{l_1})
 \bigr]+e^{-\beta E_1}\bigl[g_{0\sigma}(i\omega_l)g_{\sigma2}
 (i\omega_l) \nonumber\\
&&\quad\times\bigl(g_{0\sigma}(i\omega_{l_1})-g_{\sigma2}
 (i\omega_{l_1})\bigr)^2+
 g_{0\sigma}(i\omega_{l_1})g_{\sigma2}(i\omega_{l_1})\nonumber\\
&&\quad\times\bigl(g_{0\sigma}^2(i\omega_l)+g_{\sigma2}^2(i\omega_l)
 \bigr)\bigr]\bigr\}-Z_0^{-2}
 U^2\beta\delta_{ll_1}\bigl(e^{-\beta(E_0+E_2)} \nonumber\\
&&\quad+2e^{-\beta(E_0+E_1)}+3e^{-2\beta
 E_1}+2e^{-\beta(E_1+E_2)}\bigr)
 g_{0\sigma}^2(i\omega_l)\nonumber\\
&&\quad\times g_{\sigma2}^2(i\omega_l)+Z_0^{-2}U^2\beta
 \bigl(2e^{-\beta(E_0+E_2)}+e^{-\beta(E_0+E_1)} \nonumber\\
&&\quad+e^{-\beta(E_1+E_2)}\bigr)g_{0\sigma}(i\omega_l)g_{\sigma2}
 (i\omega_l)g_{0\sigma}(i\omega_{l_1})g_{\sigma2}(i\omega_{l_1}),
 \nonumber\\
&&\label{secondc}
\end{eqnarray}
where $g_{ij}(i\omega_l)=(i\omega_l+E_{ij})^{-1}$ is the Fourier
transform of functions~(\ref{fermion}) and~(\ref{boson}).

In this work the case of half-filling is considered. In this case
$\mu=\frac{U}{2}$ and cumulants~(\ref{firstc}) and~(\ref{secondc}) are
significantly simplified if we additionally suppose that $T\ll U$:
\begin{eqnarray}
&&K_1(i\omega_l)=\frac{i\omega_l}{(i\omega_l)^2-\bigl(\frac{U}{2}
 \bigr)^2},\nonumber\\
&&\sum_{\sigma_1}K_2(i\omega_l\sigma,i\omega_{l_1}\sigma_1,
 \omega_{l_1}\sigma_1)\nonumber\\
&&\quad=-\frac{U}{2}\Biggl(\frac{U^2}{\bigl[(i\omega_l)^2-\bigl(\frac{U}{2}
 \bigr)^2\bigr]\bigl[(i\omega_{l_1})^2-\bigl(\frac{U}{2}
 \bigr)^2\bigr]^2}\label{halffil}\\
&&\quad\quad+\frac{2[(i\omega_l)^2+\bigl(\frac{U}{2}
 \bigr)^2]}{\bigl[(i\omega_l)^2-\bigl(\frac{U}{2}
 \bigr)^2\bigr]^2\bigl[(i\omega_{l_1})^2-\bigl(\frac{U}{2}
 \bigr)^2\bigr]}\Biggr)\nonumber\\
&&\quad\quad-\beta\delta_{ll_{1}}\frac{3\bigl(\frac{U}{2}\bigr)^2}{(i
 \omega_l)^2-\bigl(\frac{U}{2}\bigr)^2}.\nonumber
\end{eqnarray}
The first term in the equation for $\sum_{\sigma_1}K_2$ is the even
function of $i\omega_{l_1}$. It does not contribute to the sum over
$l_1$ in Eq.~(\ref{approx}), since $\sum_{\bf k}t^2_{\bf k}G({\bf
k},i\omega_{l_1})$ is supposed to be an odd function of
$i\omega_{l_1}$. Thus, finally Eq.~(\ref{approx}) acquires the form
\begin{eqnarray}
K(i\omega_l)&=&\frac{i\omega_l}{(i\omega_l)^2-\bigl(\frac{U}{2}
 \bigr)^2},\nonumber\\
&+&\frac{3\bigl(\frac{U}{2}\bigr)^2}{\bigl[(i\omega_l)^2-\bigl(\frac{U}{2}
 \bigr)^2\bigr]^2}\frac{1}{N}\sum_{\bf k}t^2_{\bf k}G({\bf
 k},i\omega_{l_1}).\label{apprhalf}
\end{eqnarray}
Analogous equations were obtained in
Refs.~\onlinecite{Izyumov,Vladimir,Pairault} (in
Ref.~\onlinecite{Izyumov} the last term in the right-hand side of the
equation is three times smaller, because diagrams with spin propagators
were not taken into account there).

Let us turn to real frequencies by substituting $i\omega_l$ with
$z=\omega+i\eta$ where $\eta$ is a small positive constant which
affords an artificial broadening. Results given in the next section
were calculated with $G({\bf k},z)$ in the right-hand side of
Eq.~(\ref{apprhalf}) taken from the Hubbard-I approximation. As
mentioned, this Green's function is obtained if $K({\bf k}\omega)$ in
Eq.~(\ref{tLarkin}) is approximated by the first
cumulant~(\ref{firstc}). For half-filling we find
\begin{eqnarray}
&&G({\bf k},z)=\frac{1}{2}\left(1+\frac{t_{\bf k}}{\sqrt{U^2+t_{\bf
 k}^2}}\right)\frac{1}{z-\varepsilon_{1,\bf k}}\nonumber\\
&&\quad+\frac{1}{2}\left( 1-
 \frac{t_{\bf k}}{\sqrt{U^2+t_{\bf k}^2}}\right)
 \frac{1}{z-\varepsilon_{2,\bf k}},\nonumber\\[-1ex]
&&\label{HubbardI}\\[-1ex]
&&\varepsilon_{1,\bf k}=\frac{1}{2}\left(t_{\bf k}+\sqrt{U^2+t_{\bf
 k}^2}\right),\nonumber\\
&&\varepsilon_{2,\bf k}=\frac{1}{2}\left(t_{\bf k}-
 \sqrt{U^2+t_{\bf k}^2}\right).\nonumber
\end{eqnarray}
Below the two-dimensional square lattice is considered. It is supposed
that only the hopping constants between nearest neighbor sites $t$ are
nonzero which gives $t_{\bf k}=2t[\cos(k_x)+\cos(k_y)]$ where the
intersite distance is taken as the unit of length.

\section{Spectral function}
\begin{figure}[t]
\includegraphics[width=8.5cm]{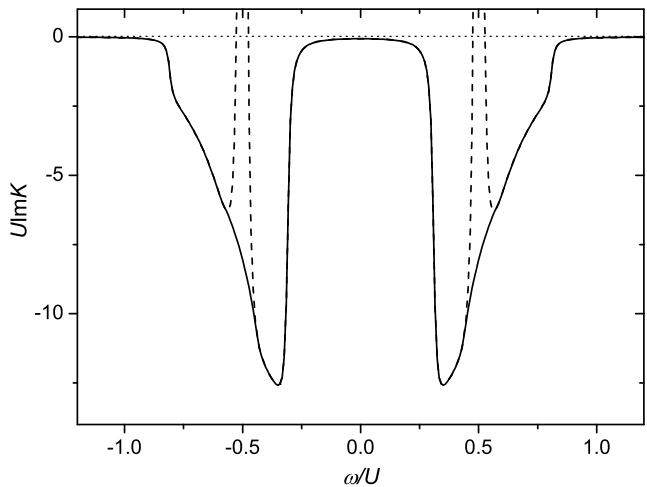}
\caption{The imaginary part of $K(\omega)$ calculated using
Eqs.~(\protect\ref{apprhalf}) and~(\protect\ref{HubbardI}) for a
100$\times$100 lattice, $t=-\frac{1}{8}U$ and $\eta=0.01U$ (the dashed
line). The solid line shows the corrected $\Im K(\omega)$ (see text).}
\label{Fig_ii}
\end{figure}

Figure~\ref{Fig_ii} demonstrates $\Im K(\omega)$ calculated with the
use of Eqs.~(\ref{apprhalf}) and~(\ref{HubbardI}). The change to real
frequencies carried out in the previous section converts the Matsubara
function~(\ref{GF}) into the retarded Green's function.
\cite{Abrikosov} It is an analytic function in the upper half-plane
which requires that $\Im K(\omega)$ be negative. As seen from
Fig.~\ref{Fig_ii}, this condition is violated at
$\omega=\pm\frac{U}{2}$. This difficulty of the considered
approximation was indicated in Ref.~\onlinecite{Pairault}. The problem
is connected with divergencies at $\omega_d=\mu$ and $\mu-U$ introduced
by functions $g_{0\sigma}(\omega)$ and $g_{\sigma 2}(\omega)$ in the
above formulas. As can be seen from the procedure of calculating the
cumulants in the previous section, these functions and divergencies
with sign-changing residues will appear in all orders of the
perturbation expansion~(\ref{series}). It can be expected that in the
entire series the divergencies compensate each other, however, in the
considered subset of terms such compensation does not occur.
Nevertheless, as seen from Fig.~\ref{Fig_ii}, at frequencies
neighboring to $\omega_d$ cumulants are regular and, if the used subset
of diagrams is expected to give a correct estimate of the entire series
for these frequencies, the value of $\Im K(\omega)$ at the singular
frequencies can be reconstructed using an interpolation and its values
in the regular region. An example of such interpolation is given in
Fig.~\ref{Fig_ii}.

The function $K(z)$ has to be analytic in the upper half-plane also and
therefore its real part can be calculated from its imaginary part using
the Kramers-Kronig relations. We use this way with the interpolated
$\Im K(\omega)$ to avoid the influence of the divergencies on $\Re
K(\omega)$. However, the use of the interpolation overrates somewhat
values of $|\Im K(\omega)|$ which leads to the overestimation of the
tails in the real part. To correct this defect the interpolated
$K(\omega)$ is scaled so that its real part for frequencies
$|\omega|\gg\frac{U}{2}$ coincide with the values obtained from
Eq.~(\ref{apprhalf}). It is worth noting that such obtained spectral
function
\begin{eqnarray}
A({\bf k}\omega)&=&-\frac{1}{\pi}\Im G({\bf k}\omega)\nonumber\\
&=&-\frac{1}{\pi} \frac{\Im K(\omega)}{[1-t_{\bf k}\Re
 K(\omega)]^2+[t_{\bf k}\Im K(\omega)]^2}\label{specfun}
\end{eqnarray}
satisfies the normalization condition $\int_{-\infty}^\infty d\omega
A({\bf k}\omega)=1$ with an accuracy better than 0.01.

\begin{figure}
\includegraphics[width=8cm]{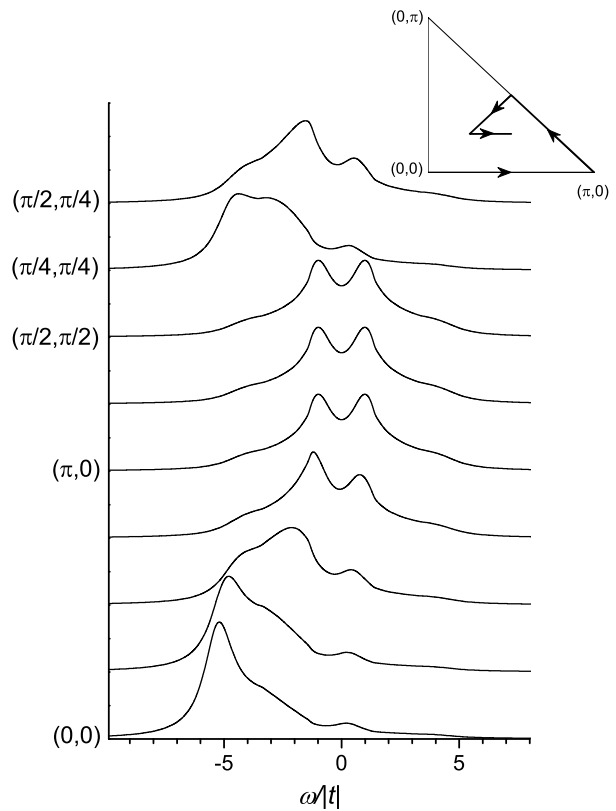}
\caption{The spectral function $A({\bf k}\omega)$ calculated for
momenta along the route shown in the inset. An 8$\times$8 lattice,
$t=-\frac{1}{4}U$ and $\eta=0.2U$.} \label{Fig_iii}
\end{figure}
The spectral function obtained in this way with the use of
Eqs.~(\ref{apprhalf}), (\ref{HubbardI}), and~(\ref{specfun}) for
momenta in the Brillouin zone of an 8$\times$8 lattice is shown in
Fig.~\ref{Fig_iii}. In these calculations the small lattice and a
comparatively large artificial broadening were used to reproduce the
conditions of Monte-Carlo calculations carried out in
Ref.~\onlinecite{Moukouri}. Comparing Fig.~\ref{Fig_iii} with Fig.~5 in
that work we find close similarity of these two groups of spectra --
not only the general shapes of spectra are close but also the position
of maxima in them. Some discrepancies can be also noticed. In
particular, in the spectrum for ${\bf k}=(\frac{\pi}{2},\frac{\pi}{4})$
in Fig.~\ref{Fig_iii} the intensity of maximum at a positive frequency
is larger than in the respective spectrum in Ref.~\onlinecite{Moukouri}
(in this latter spectrum this maximum looks like a barely perceptible
peculiarity).

\begin{figure}
\includegraphics[width=8cm]{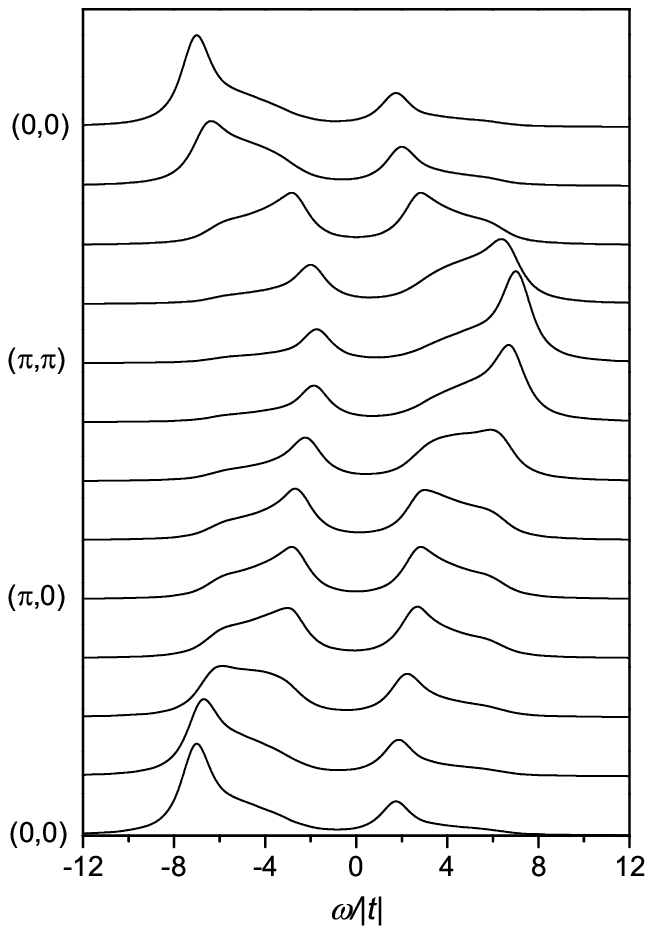}

\vspace{5ex}
\includegraphics[width=8cm]{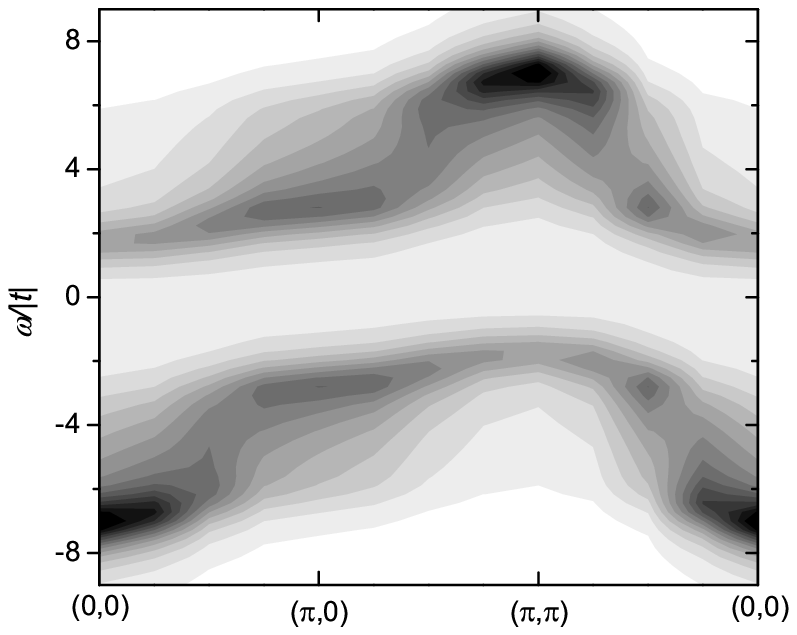} \caption{The upper panel: the
spectral function $A({\bf k}\omega)$ calculated for momenta along the
symmetry lines of the square Brillouin zone. The lower panel: the
dispersion of maxima in the upper panel shown as a gray-scale plot with
darker areas corresponding to larger intensity of maxima. A
100$\times$100 lattice, $t=-\frac{1}{8}U$ and $\eta=0.1U$.}
\label{Fig_iv}
\end{figure}

The spectral function for a larger $U$ is shown in Fig.~\ref{Fig_iv}
where the momenta along the symmetry lines of the square Brillouin zone
were chosen. Let us compare these results with the spectra in Fig.~1 in
Ref.~\onlinecite{Grober} which were obtained in Monte-Carlo
calculations with the same ratio $U/|t|$ in an 8$\times$8 lattice. We
found that the results shown in Fig.~\ref{Fig_iv} change only slightly
with decreasing the lattice size to this value. The comparison has to
be carried out with low-temperature spectra of
Ref.~\onlinecite{Grober}, since Eq.~(\ref{halffil}) used in our
calculations was derived for the case $T\ll U$. Notice that in this
limit at half-filling the spectral function in the used approximation
does not depend on temperature. Again we find that our calculated
spectra are close to those obtained in the Monte-Carlo calculations,
though there are some differences in relative intensity of maxima. In
our spectra some maxima resolved in the Monte-Carlo calculations look
like shoulders of main maxima and a decrease of the artificial
broadening $\eta$ does not convert the shoulders to maxima. On the
other hand, the maxima at $\omega\approx 2|t|$ in the spectra for
momenta near $(0,0)$ and at $\omega\approx -2|t|$ in the spectra for
momenta near $(\pi,\pi)$ which are well resolved in our results are
barely perceptible in Fig.~1 of Ref.~\onlinecite{Grober}. The possible
reason for these differences can be established from the comparison of
Figs.~\ref{Fig_iii} and~\ref{Fig_iv}. The respective maxima in the
former figure for the smaller $U$ are less intensive. Apparently, like
the Hubbard-I approximation, the used approach somewhat overestimates
the electron correlations. Nevertheless, Fig.~\ref{Fig_iv} reproduces
correctly the most essential features of the spectra obtained by the
Monte-Carlo method.

One of these features recently reproduced also in the cluster
perturbation \cite{Dahnken} and the two-particle self-consistent
\cite{Tremblay} calculations is the four-band structure of the
spectrum. In Fig.~\ref{Fig_iv} these bands are located near frequencies
$\omega=\pm 3|t|$ and $\omega=\pm 6|t|$. The high-frequency bands are
observed near the $(0,0)$ and $(\pi,\pi)$ points, while the
low-frequency bands have a larger existence domain in our calculations.
Similar four bands can be identifies in Fig.~\ref{Fig_iii}. A more
detailed treatment of the low-frequency bands shows that either of them
in turn contain two components. Maxima in $A({\bf k}\omega)$ arise by
two reasons -- due to the smallness of the denominator in
Eq.~(\ref{specfun}) when a frequency satisfying the equation $1-t_{\bf
k}\Re K(\omega)=0$ falls into a region of a small damping $|\Im
K(\omega)|$ and due to the strong frequency dependence of $\Im
K(\omega)$ in the numerator of Eq.~(\ref{specfun}). The high-frequency
bands and the low-frequency bands in the central part and at the
periphery of the Brillouin zone arise due to the first reason. As seen
from Fig.~\ref{Fig_ii}, the regions of small damping are located on
either side of the two maxima in $\Im K(\omega)$ which gives rise to
the well-separated four bands. However, for momenta which are close to
the boundary of the magnetic Brillouin zone the equation $1-t_{\bf
k}\Re K(\omega)=0$ has no solutions due to a small or vanishing $t_{\bf
k}$. In this case the two maxima of $A({\bf k}\omega)$ originate from
the two maxima in $|\Im K(\omega)|$. They are located in the same
frequency ranges as the low-frequency maxima originated from the
resonance denominator in Eq.~(\ref{specfun}). Therefore they can be
combined in a common dispersion band, although widths of the maxima of
different origin can significantly vary for a smaller artificial
broadening.

As can be seen from Eqs.~(\ref{tLarkin}), (\ref{apprhalf}),
and~(\ref{HubbardI}), the used approximation does not describe the Mott
transition -- for $\eta\rightarrow 0$ the density of states
$\rho(\omega)=N^{-1}\sum_{\bf k}A({\bf k}\omega)$ at $\omega=0$
vanishes for any finite $U$. This flaw can be remedied by going to the
next approximation. The use of approximation~(\ref{HubbardI}) in
Eq.~(\ref{apprhalf}) means that only diagrams (a) in Fig.~\ref{Fig_i}
are inserted in the hopping line of the diagram (b). If these latter
diagrams are also inserted in the line, Eq.~(\ref{apprhalf}) is
transformed to the following self-consistent equation for the
irreducible part:
\begin{eqnarray}
K(z)&=&\frac{z}{z^2-\bigl(\frac{U}{2}\bigr)^2},\nonumber\\
&+&\frac{3\bigl(\frac{U}{2}\bigr)^2}{\bigl[z^2-\bigl(\frac{U}{2}
 \bigr)^2\bigr]^2}\frac{1}{N}\sum_{\bf k}\frac{t_{\bf k}}{1-t_{\bf
 k}K(z)}. \label{selfcons}
\end{eqnarray}
For the initial electron dispersion $t_{\bf k}$ described by the
semielliptical density of states
$$\rho_0(\omega)=\left\{\begin{array}{rl}\frac{2}{\pi B}\sqrt{1-
 \left(\frac{\omega}{B}\right)^2}, & |\omega|\leq B, \\ 0, &
 |\omega|> B,\end{array} \right. $$
where $B$ is the halfwidth of the band, Eq.~(\ref{selfcons}) can be
solved analytically and the condition for the disappearance of the gap
at $\omega=0$ in $\rho(\omega)$ can be found. It happens when $U\leq
U_c=\sqrt{3}B$ and the gap opens when $U>U_c$. Analogous results for
somewhat different equations and initial dispersions were obtained in
Refs.~\onlinecite{Zaitsev,Izyumov,Pairault}.

\section{Conclusion}
The considered diagram technique is very promising for a generalization
to many-band Hubbard models for which energy parameters of the one-site
parts of the Hamiltonians exceed or at least are comparable to the
intersite parameters. The expansion in powers of these latter
parameters can be expressed in terms of cumulants in the same manner as
discussed above. Now there are distinct cumulants which belong to
different site states characterized by dissimilar parameters of the
repulsion and the level energy. These cumulants are described by
formulas similar to Eqs.~(\ref{firstc}) and~(\ref{secondc}). For
example, for the Emery model \cite{Emery} which describes oxygen
$2p_\sigma$ and copper $3d_{x^2-y^2}$ orbitals of Cu-O planes in
high-$T_c$ superconductors there are two types of cumulants
corresponding to these states. Diagrams of the lowest orders, e.g., for
Green's function on copper sites resemble those shown in
Fig.~\ref{Fig_i} where ``oxygen'' cumulants are included in hopping
lines. Equations of the type of Eq.~(\ref{tLarkin}) and partial
summations similar to Eq.~(\ref{thopping}) can be derived also in this
case.

In summary, the diagram technique for the one-band Hubbard model was
formulated for the case of moderate to strong Hubbard repulsion. The
expansion in powers of the hopping constant is expressed in terms of
site cumulants of electron creation and annihilation operators. For
Green's function the equation of the Larkin type was derived and solved
for the case of two dimensions, nearest-neighbor hopping and
half-filling. The obtained spectrum contains four bands. It was shown
that maxima of the spectral function forming the bands are of
dissimilar origin in different regions of the Brillouin zone -- in its
central part and at the periphery the maxima stem from the resonance
denominator in the formula for the spectral function, while near the
boundary of the magnetic Brillouin zone they originate from the strong
frequency dependence of the irreducible part in the numerator of this
formula. The obtained band structure and the shapes of the spectral
function are close to those found in the Monte Carlo calculations.

\begin{acknowledgments}
This work was supported by the ESF grant No.~5548.
\end{acknowledgments}

\end{document}